\documentclass[10pt,conference]{IEEEtran}

\usepackage[pdftex]{graphicx}
\usepackage{multirow}
\usepackage[dvipsnames]{xcolor}
\usepackage{cite}
\usepackage{subcaption}

\newcommand{\libAtk}{\texttt{Library-Attack}}
  
\begin{document}

\title{\textbf{\libAtk}: Reverse Engineering Approach for Evaluating Hardware IP Protection\\
}

\author{
\IEEEauthorblockN{Aritra Dasgupta, Sudipta Paria, Swarup Bhunia}
\IEEEauthorblockA{Electrical and Computer Engineering\\
University of Florida\\
Gainesville, FL 32608, USA\\
Email: \{aritradasgupta, sudiptaparia\}@ufl.edu, swarup@ece.ufl.edu}
\and
\IEEEauthorblockN{Christopher Sozio, Andrew Lukefahr}
\IEEEauthorblockA{Intelligent Systems Engineering\\
Indiana University\\
Bloomington, IN 47405, USA\\
Email: \{cmsozio, lukefahr\}@iu.edu}\\
}

\maketitle

\begin{abstract}
Existing countermeasures for hardware IP protection, such as obfuscation, camouflaging, and redaction, aim to defend against confidentiality and integrity attacks. However, within the current threat model, these techniques overlook the potential risks posed by a highly skilled adversary with privileged access to the IC supply chain, who may be familiar with critical IP blocks and the countermeasures implemented in the design. To address this scenario, we introduce \libAtk, a novel reverse engineering technique that leverages privileged design information and prior knowledge of security countermeasures to recover sensitive hardware IP. During \libAtk, a privileged attacker uses known design features to curate a design library of candidate IPs and employs structural comparison metrics from commercial EDA tools to identify the closest match. We evaluate \libAtk~on transformed ISCAS89 benchmarks to demonstrate potential vulnerabilities in existing IP-level countermeasures and propose an updated threat model to incorporate them.\\

\end{abstract}
\renewcommand\IEEEkeywordsname{Keywords}

\section{Introduction}

\begin{figure}[!htbp]
    \centering
    \includegraphics[width=\columnwidth]{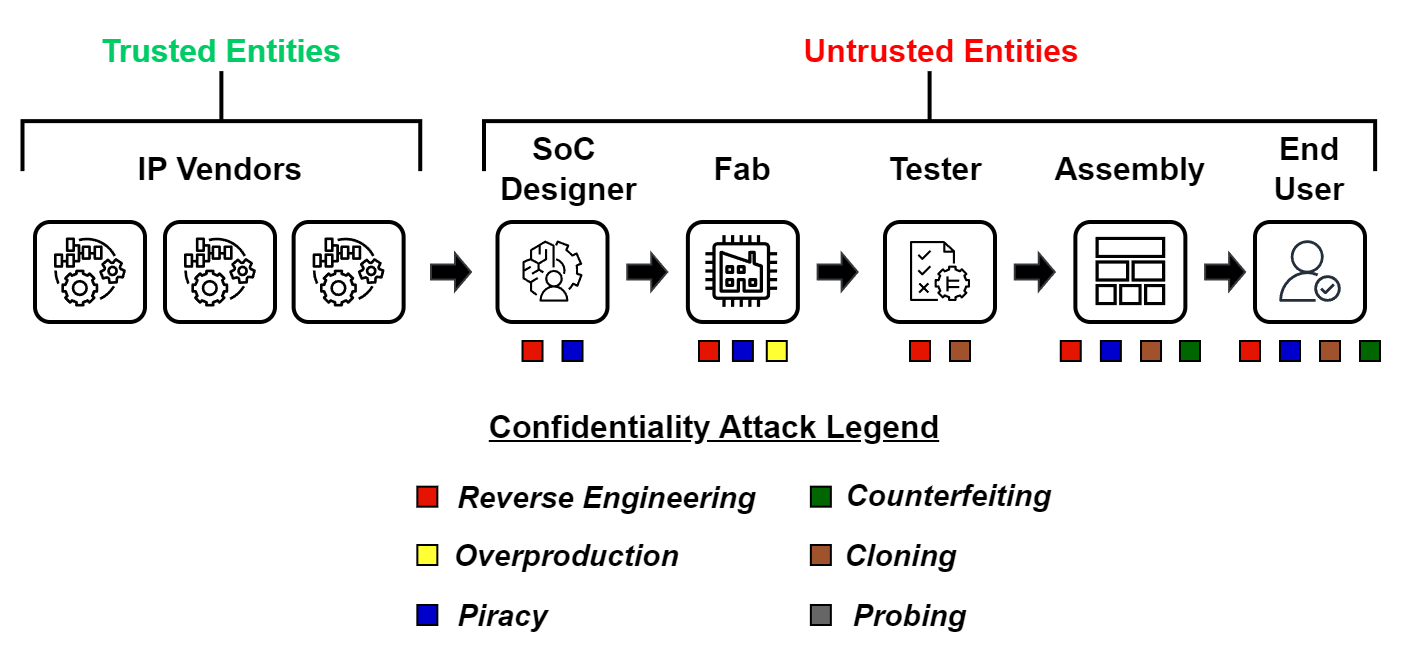}
    \caption{Various threats encountered in the IC design flow.}
    \label{fig:ic_design_flow}
\end{figure}

\begin{figure}[!htbp]
\centering
\subfloat[Countermeasures Taxonomy.]{\includegraphics[width=0.65\columnwidth]{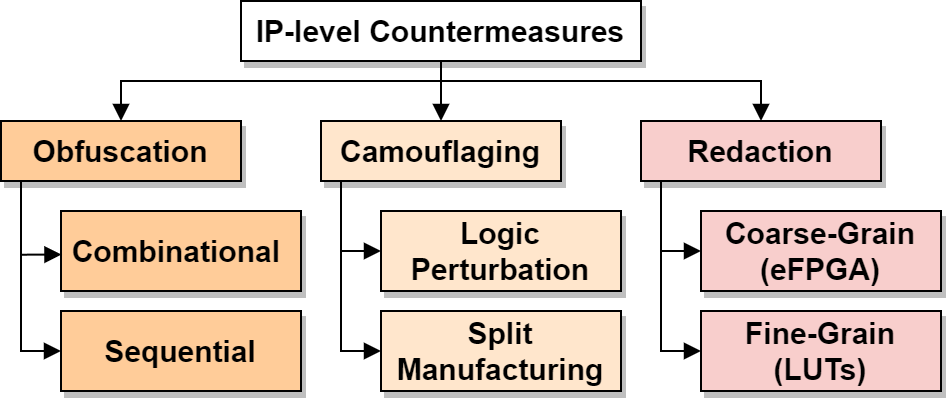}
\label{fig:taxo_counter}}
\hfil
\subfloat[Attacks Taxonomy.]{\includegraphics[width=0.9\columnwidth]{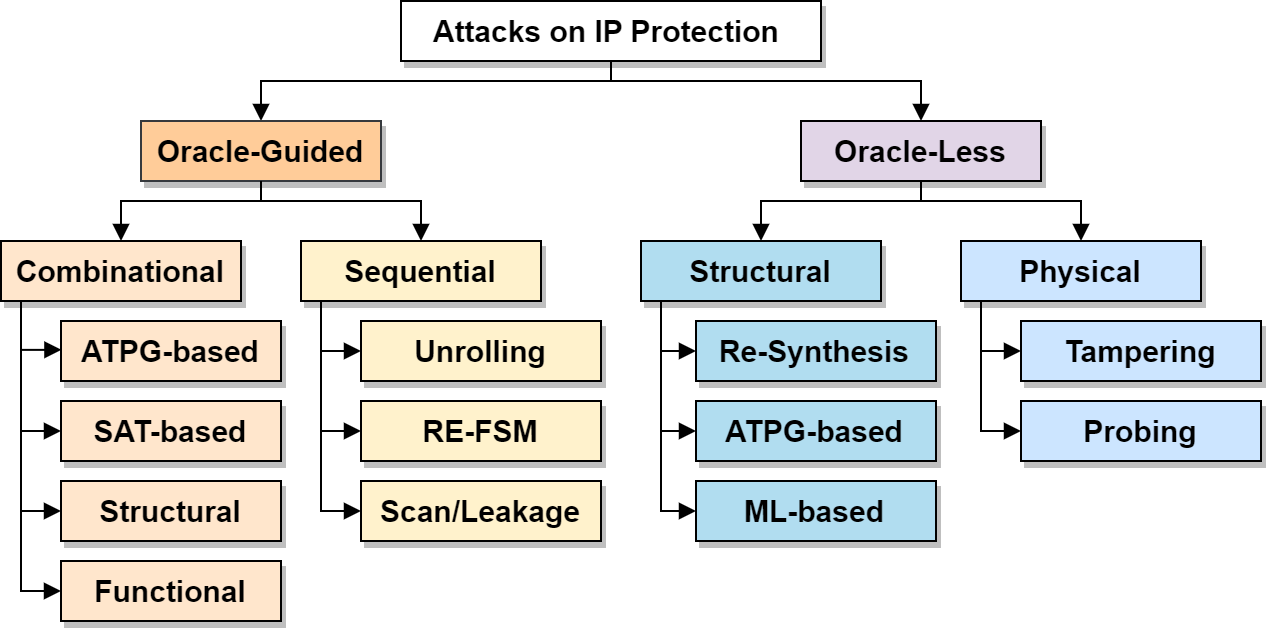}
\label{fig:taxo_attack}}
\caption{Taxonomies: (a) Existing IP-level countermeasures, and (b) various oracle-guided and oracle-less attack vectors in literature.}%
\label{fig:taxonomy}%
\vspace{-2em}
\end{figure}

The modern semiconductor industry heavily relies on third-party hardware intellectual property (IP) vendors and offshore fabrication facilities to meet the high demands for integrated circuits (IC) and reduce costs, which has led to an increase in IP piracy, counterfeiting and reverse engineering (RE) efforts by malicious entities in the supply chain \cite{hw_security_book_2018_Bhunia_Tehranipoor}, as depicted in Fig. \ref{fig:ic_design_flow}. Countermeasures proposed over the years to protect hardware IP against these threats can be broadly classified as shown in Fig. \ref{fig:taxo_counter}.
However, these techniques are shown to be vulnerable to attacks that can predict the unlocking key input sequence or bypass the countermeasure altogether \cite{ll_review_2022_Kamali}. The current threat model for IP protection categorizes these attacks based on whether the attacker has access to a functional golden design (oracle): (1) \textit{oracle-guided} attacks \cite{sat_2015_Subramanyan,appSAT_2017_Shamsi}, and (2) \textit{oracle-less}  attacks \cite{sail_2018_Chakraborty,sweep_2019_Alaql,gnnunlock_2021_Alrahis}, as shown in Fig. \ref{fig:taxo_attack}.

Recent countermeasures \cite{saro_2021_Alaql,unsail_2021_Alrahis,alice_2022_Tomajoli,practical_2023_Rahman} have tried to address these attack vectors and use them to quantify the level of assurance achieved by injecting them into a design. However, they overlook the possibility that an adversary with highly specialized expertise and authorized access to the supply chain can analyze the protected hardware IP to identify unprotected or open-source counterparts with equivalent functionality. Moreover, such an adversary can access the protection algorithm used (to satisfy Kerckhoff's principle\footnote{https://www.crypto-it.net/eng/theory/kerckhoffs.html}). In this paper, we propose \libAtk~that aims to exploit this gap in the current threat model. The proposed attack uses functional I/O features extracted from a transformed design with countermeasures to create a library of designs and then uses a similarity score based on cut-point matching to recover the original design. The major contributions in this paper are listed below:
\begin{itemize}
    \item We propose \libAtk, a novel RE methodology that leverages privileged information about a protected design and prior knowledge of the security countermeasure used to recover the original unprotected design.
    \item We demonstrate the efficacy of \libAtk~using case studies on two distinct and well-established IP-level countermeasures applied to open-source benchmarks.
    \item We propose an updated threat model for hardware IP protection that incorporates the potential risks posed by a highly skilled adversary with privileged access to the IC supply chain, highlighted by \libAtk.
\end{itemize}

The remainder of this paper is organized as follows: In Section II, we introduce the proposed \libAtk~methodology and the major steps involved. In Section III, we describe our experimental setup, present two case studies on existing IP-level countermeasures, and propose an updated threat model. Section IV concludes the paper. 

\section{\libAtk~Methodology}

\begin{figure}[!htbp]
    \centering
    \includegraphics[width=0.95\columnwidth]{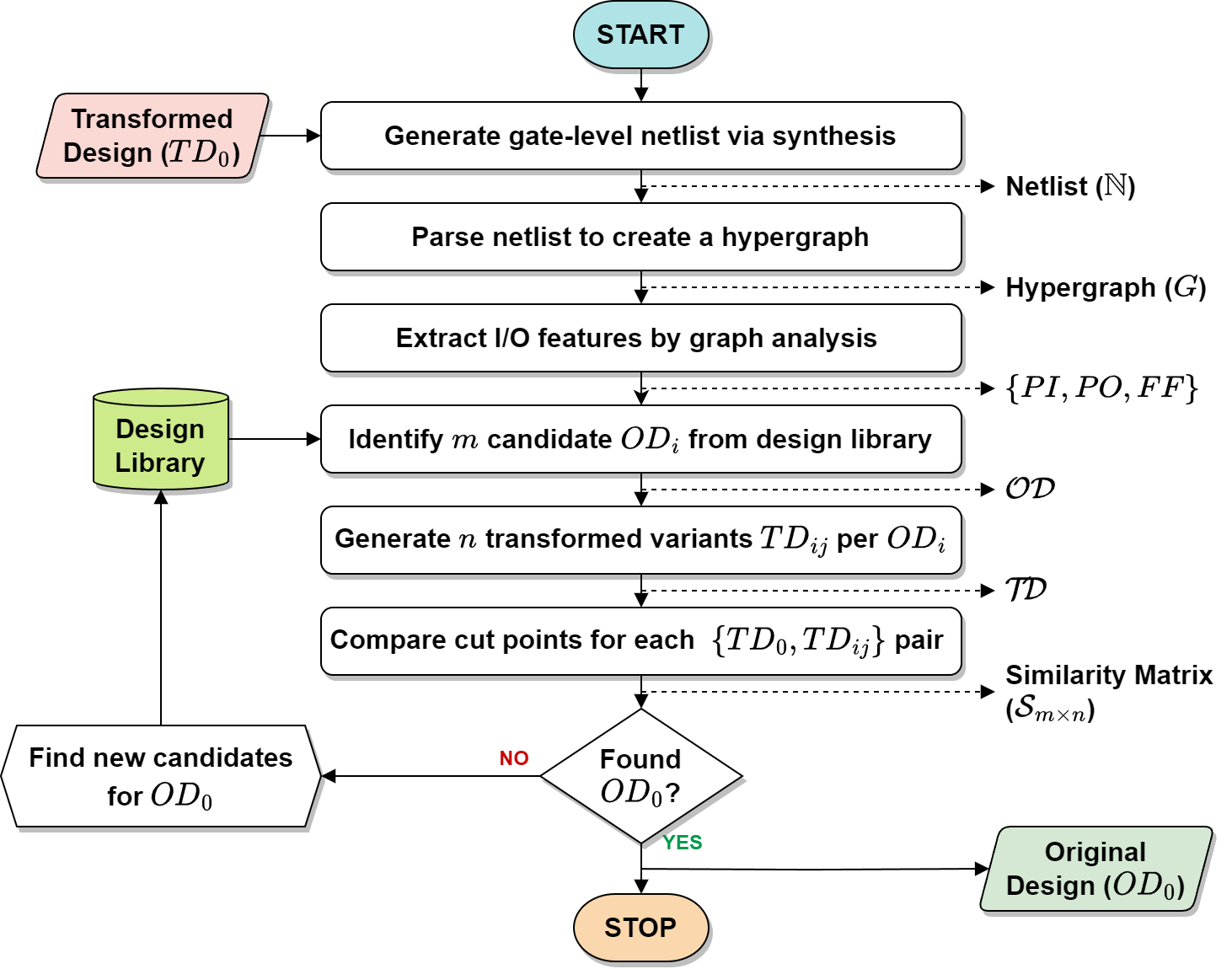}
    \caption{\libAtk~steps.}
    \label{fig:flow_libAtk}
    % \vspace{-2em}
\end{figure}

\begin{figure}[!htbp]
\centering
\includegraphics[width=0.9\columnwidth]{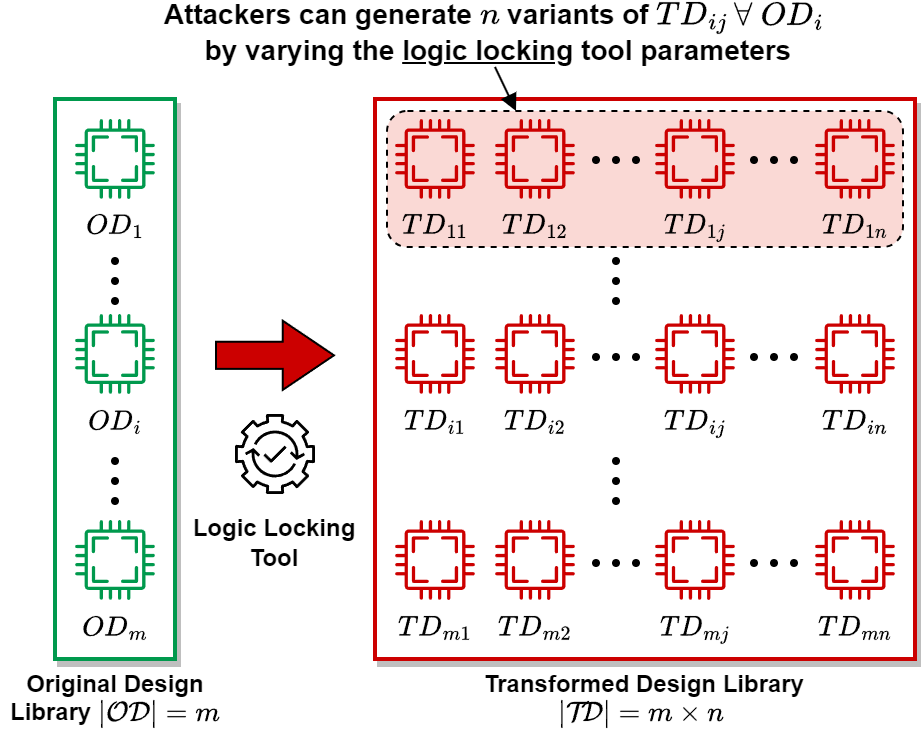}
\caption{Overview of the original and transformed design library generation in \libAtk. The attacker creates a library of $m$ candidate $OD_i$ after analyzing the functional and scan I/O ports. Next, they use the logic locking tool to generate $n$ variants for every candidate $OD_i$, resulting in a $TD_{ij}$ library of size $m \times n$.}
\label{fig:libAtk_overview}
\end{figure}

Fig. \ref{fig:flow_libAtk} shows the different steps involved in \libAtk. We assume that the attacker has access to the protected transformed design ($TD_0$) and the countermeasure used. The gate-level netlist of $TD_0$ is converted to a hypergraph ($G$) consisting of logic gates as vertices connected by edges derived from wires. The attacker analyzes $G$ to extract the I/O features, namely the primary inputs ($PI$), primary outputs ($PO$), and the flip-flops ($FF$). Using these features, the attacker can identify a set of $m$ candidates for the original design ($\mathcal{OD}$) from a library of known designs and other open-source benchmarks. The attacker then varies the configurable parameters the \textit{logic locking tool} to generate $n$ transformed design variants $\forall ~OD_i \in \mathcal{OD}$, resulting in a transformed design library $\mathcal{TD}$ of size $m \times n$, as shown in Fig. \ref{fig:libAtk_overview}. The cut-points ($PO$ and $FF$) from each pair of $\{TD_0,TD_{ij}\}$ ($1~\leq i~\leq~m,1~\leq~j~\leq~n$) are compared using a structural analysis tool that can match similar cut-points and generates an overall similarity score between $TD_0$ and $TD_{ij}$, which can vary from 0 (no matching cut-points) and 1 (all cut-points are identical). The similarity scores are consolidated in the form of a similarity score matrix $\mathcal{S}_{m \times n}$, and the ${OD_i}$ with the highest overall score across the variants is the original design $OD_0$ used to generate $TD_0$. If $OD_0$ cannot be recovered from $\mathcal{S}_{m \times n}$, or in the case of multiple candidates with identical highest score, the design library needs to be updated with new potential candidates for $OD_0$.

\section{Results and Analysis}

\begin{figure*}[!htbp]
\centering
\subfloat[Original Netlist.]{\includegraphics[width=0.3\textwidth]{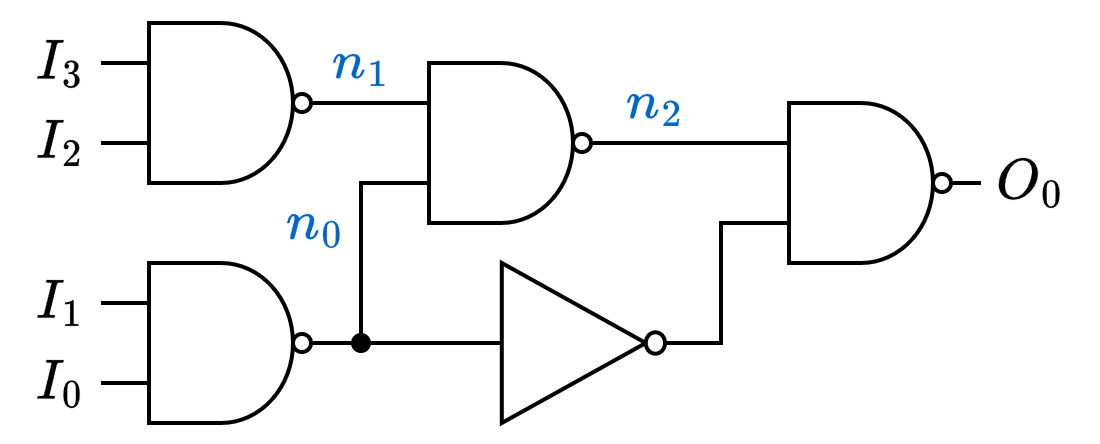}
\label{fig:ll_org}}
\hfil
\subfloat[XOR Locked Netlist.]{\includegraphics[width=0.36\textwidth]{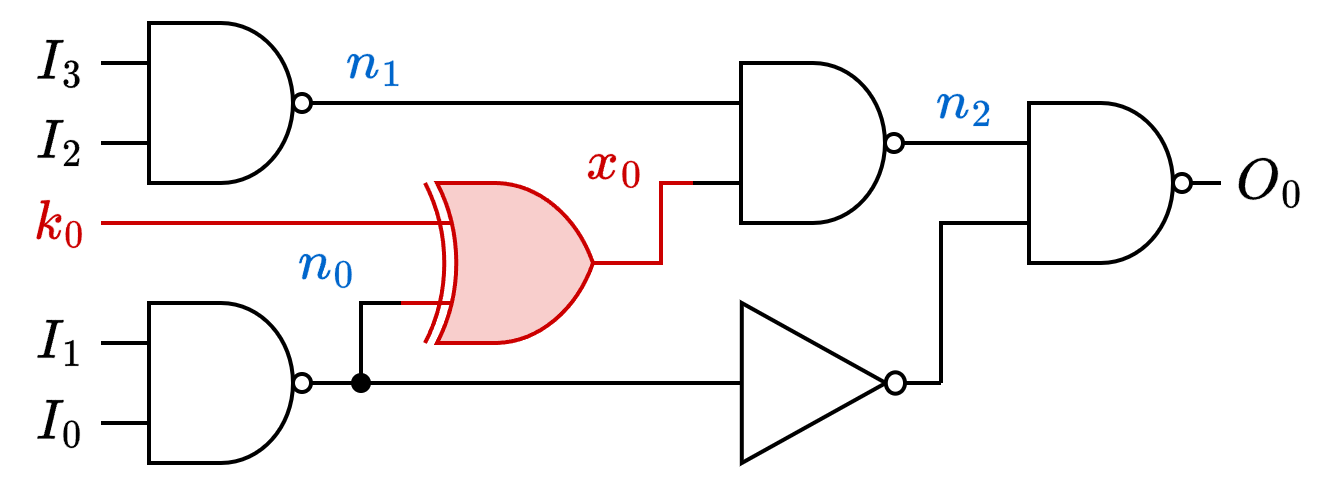}
\label{fig:ll_xor}}
\hfil
\subfloat[LUT Obfuscated Netlist.]{\includegraphics[width=0.31\textwidth]{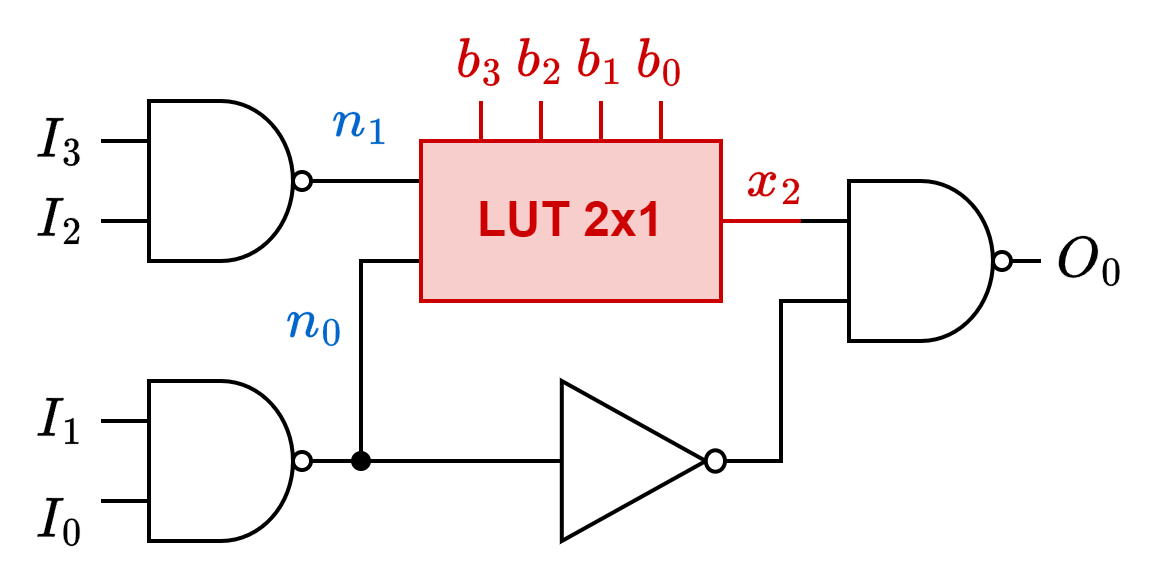}
\label{fig:ll_lut}}
\caption{Overview IP-level countermeasures: (a) Original gate-level netlist; (b) XOR locking using a key-gate, the true logic at $n_0$ is only restored when the keyinput $k_0$ is set to 0; and (c) LUT obfuscation where a NAND2 gate is replaced with a LUT 2x1, the true functionality at $n_2$ is restored only when the bitstream $\{b_3,b_2,b_1,b_0\}$ is configured to 4'b0111.}%
\label{fig:ll_demo}%
\end{figure*}

\subsection{Experimental Setup}
We evaluated the proposed \libAtk~on two well-known IP-level countermeasures: (1) 128-bit XOR Locking \cite{epic_2008_roy} and (2) 128-bit LUT Obfuscation \cite{LUTLock_2018_Kamali} (using 8 LUT4x1). Fig. \ref{fig:ll_demo} demonstrates the transformations introduced by these countermeasures using a small gate-level netlist as example. For our analysis, we chose the $s298$ and $s526$ traffic light controllers from the open-source ISCAS89 benchmark suite\footnote{https://ddd.fit.cvut.cz/www/prj/Benchmarks/} as the designs under test ($OD_0$). Of the 40 ISCAS89 benchmarks, Table \ref{tab:od_library} lists the 5 designs with similar functionality as $s298$ and $s526$ and identical counts of $PI$ and $PO$, which constitute the original design library $\mathcal{OD}$ for \libAtk. For both countermeasures, the transformed designs under test ($s298$ and $s526$) and all variants comprising the transformed design library $\mathcal{TD}$ were generated using the NEOS tool\footnote{https://bitbucket.org/kavehshm/neos/src/master/}. The evaluation designs from both $\mathcal{OD}$ and $\mathcal{TD}$ libraries were mapped to the NanGate 45nm open standard cell library\footnote{https://si2.org/open-cell-and-free-pdk-libraries/} using Synopsys Design Compiler (V-2023.12-SP5). Cadence Conformal (21.10-s300) is used to perform Logic Equivalence Checking (LEC), and the design similarity scores reported by LEC were used for the cut-point comparison step in \libAtk. All experiments were carried out on a Red Hat Enterprise Linux Server 7.9 server with AMD® Epyc 7713 64-core processor and 1007.6 GiB memory.

\begin{table}[!htbp]
\centering
\caption{ISCAS89 benchmarks (traffic light controllers) that constitute the $\mathcal{OD}$ with $\left|PI\right| = 3$ and $\left|PO\right| = 6$.}
\label{tab:od_library}
\resizebox{0.8\columnwidth}{!}{
\begin{tabular}{ccccc}
\hline
\textbf{Benchmark} & \textbf{\#PI} & \textbf{\#PO} & \textbf{\#FF} & \textbf{\#Gates} \\
\hline
\textbf{s298}   & 3 & 6 & 14 & 119 \\
\textbf{s382}   & 3 & 6 & 21 & 162 \\
\textbf{s400}   & 3 & 6 & 21 & 158 \\
\textbf{s444}   & 3 & 6 & 21 & 181 \\
\textbf{s526}   & 3 & 6 & 21 & 193 \\
\hline
\end{tabular}
}
\end{table}

\subsection{Case Studies}

\begin{figure}[!htbp]
\centering
\subfloat[]{\includegraphics[width=0.75\columnwidth]{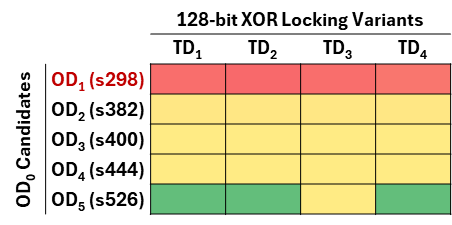}
\label{fig:s298_xor}}
\hfil
\subfloat[]{\includegraphics[width=0.75\columnwidth]{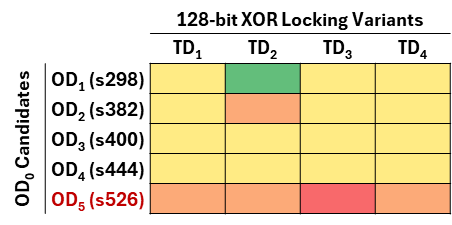}
\label{fig:s526_xor}}
\caption{Heatmaps representing LEC similarity matrices generated during \libAtk~on 128-bit XOR Locking for: (a) $s298$ as $OD_0$, and (b) $s526$ as $OD_0$. In both cases, the heatmaps demonstrate that the $OD_0$ was correctly recovered out of the library with ($\left|\mathcal{OD}\right|$ $=5$) candidates using the normalized LEC similarity scores (ranges from \textcolor{Green}{0} to \textcolor{Red}{1}).}%
\label{fig:lec_sim_xor}%
% \vspace{-1em}
\end{figure}

\noindent\textbf{1. 128-bit XOR Locking:} \\
We evaluated \libAtk~on the $s298$ and $s526$ ISCAS89 benchmarks, where each benchmark was transformed by randomly placing 128 XOR/XNOR key gates at suitable locations, resulting in a 128-bit unlocking key. The true functionality of the transformed design is restored only when the correct 128-bit key value is applied. For each candidate ${OD_i} \in \mathcal{OD}$, 4 different variants were generated using the NEOS tool by varying the seed of the built-in random number generator. The heatmaps depicted in Fig. \ref{fig:lec_sim_xor} represent the normalized LEC similarity matrices generated during \libAtk, and the similarity scores can vary between \textcolor{Green}{0} (lowest similarity) to \textcolor{Red}{1} (highest similarity). For both transformed designs under test, ${OD_0}$ was successfully recovered from the 5 candidates in $\mathcal{OD}$, as observed from the heatmaps in Fig. \ref{fig:lec_sim_xor}.

\begin{figure}[!htbp]
\centering
\subfloat[]{\includegraphics[width=0.75\columnwidth]{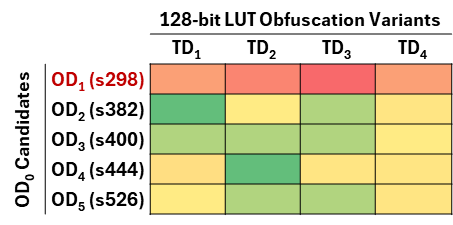}
\label{fig:s298_lut}}
\hfil
\subfloat[]{\includegraphics[width=0.75\columnwidth]{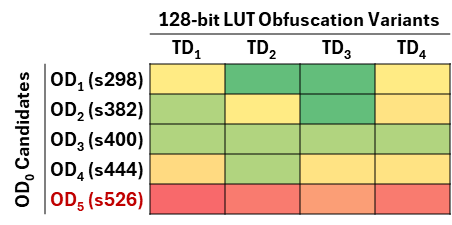}
\label{fig:s526_lut}}
\caption{Heatmaps representing LEC similarity matrices generated during \libAtk~on 128-bit LUT Obfuscation for: (a) $s298$ as $OD_0$, and (b) $s526$ as $OD_0$. In both cases, the heatmaps demonstrate that the $OD_0$ was correctly recovered out of the library with ($\left|\mathcal{OD}\right|$ $=5$) candidates using the normalized LEC similarity scores (ranges from \textcolor{Green}{0} to \textcolor{Red}{1}).}%
\label{fig:lec_sim_lut}%
\end{figure}

\noindent\textbf{2. 128-bit LUT Obfuscation:} \\
We use \libAtk~to evaluate 128-bit LUT Obfuscation on the same $s298$ and $s526$ benchmarks from ISCAS89, and the NEOS tool is used to transform the designs by randomly inserting 8 configurable LUT4x1 cells. All LUTs in the obfuscated designs need to be configured using the correct bitstream sequence to restore the true functionality. In order to generate 4 different variants for each candidate ${OD_i} \in \mathcal{OD}$, the NEOS tool randomly identifies the logic cones to be replaced with LUTs. Fig. \ref{fig:lec_sim_lut} shows the heatmaps generated during \libAtk, and it can be observed that ${OD_0}$ was successfully recovered for both transformed designs under test.

\subsection{Discussion}
It should be noted that the two countermeasures studied were already broken by existing attack vectors (shown in Fig. \ref{fig:taxo_attack}) such as \cite{sat_2015_Subramanyan,sail_2018_Chakraborty,sweep_2019_Alaql,gnnunlock_2021_Alrahis}, and the analyzed benchmarks were relatively small, which somewhat limited the structural variance possible in the transformed designs. However, the case studies clearly demonstrate that \libAtk~can successfully recover the original unprotected benchmarks regardless of the nature of the transformations performed. Furthermore, \libAtk~can be combined with existing techniques such as SAT \cite{sat_2015_Subramanyan} or SWEEP \cite{sweep_2019_Alaql} to strengthen these attacks and increase the probability of breaking the employed countermeasures.

Motivated by our observations from \libAtk, we propose an \textbf{updated threat model for hardware IP protection}:

\begin{itemize}
    \item \textbf{Assets:} Gate-level hardware IPs (with associated design files), transformed and protected using IP-level countermeasures such as logic locking or obfuscation.
    \item \textbf{Adversary:} Highly skilled entities from untrusted third-party facilities (fabrication/testing) and untrusted users (post-deployment); with access to privileged information about sensitive IPs (and the countermeasures used) that can be leveraged to generate a library of candidate IPs with similar structural/functional features as the original IP. They are also familiar with commercial EDA tools and licensed software suitable for RE.
    \item \textbf{Adversarial objectives:} To recover the unprotected original IP from a library of candidate IPs and/or extract design secrets.
    \item \textbf{Trust Model:} The IP owner/designer is considered trustworthy.
\end{itemize}

\section{Conclusion}
In this paper, we presented \libAtk, a novel RE technique that leverages privileged design information and prior knowledge of IP-level countermeasures to recover the original unprotected IP from a library of candidate IPs similar to the original. We described the methodology of the proposed attack and subsequently evaluated \libAtk~on ISCAS89 benchmarks transformed using two different IP-level countermeasures. We demonstrated that \libAtk~can successfully identify and recover the original unprotected IPs from a library of candidate ISCAS89 benchmarks, regardless of the transformation or the benchmark under test. Finally, we proposed an updated threat model for hardware IP protection incorporating the potential vulnerabilities exposed by \libAtk~technique. 

\bibliographystyle{IEEEtran}
\bibliography{references}

% that's all folks
\end{document}